\documentclass[sigconf]{sig-alternate}

\usepackage[ruled,vlined]{algorithm2e}
\usepackage{epsfig}
\usepackage{epstopdf}
\usepackage{url}
\usepackage{cite}
\usepackage{subfigure}
\usepackage{authblk}
\usepackage[skip=5pt]{caption}
\usepackage{color}
\usepackage{amsmath}
\providecommand{\eat}[1]{}

\makeatletter
\def\@copyrightspace{\relax}
\makeatother

\begin{document}
\title{Realization of CDMA-based IoT Services with Shared Band Operation of LTE in 5G}

\author[1]{Shweta S. Sagari}
\author[2]{Siddarth Mathur}
\author[3]{Dola Saha}
\author[1]{Syed Obaid Amin}
\author[1]{Ravishankar Ravindran}
\author[2]{Ivan Seskar}
\author[2]{\\Dipankar Raychaudhuri}
\author[1]{Guoqiang Wang \vspace*{-0.2cm}}

\affil[1]{Huawei Research Center, CA, {\em\{shweta.sagari, obaid.amin, ravi.ravindran, gq.wang\}@huawei.com}}
\affil[2]{WINLAB, Rutgers University, NJ, 
{\em\{siddarthmathur, seskar, ray\}@winlab.rutgers.edu}}
\affil[3]{University at Albany, SUNY, NY, {\em dsaha@albany.edu}}

\eat{
\author[1]{A1}
\author[1]{A2}
\author[2]{A3}
\author[3]{A4}
\affil[1]{Affiliation1, {\em\{a1, a2\}@email.com}}
\affil[2]{Affiliation2, {\em  a3@email.com}}
\affil[3]{Affiliation2, {\em a4@email.com}}
}

\maketitle

\eat{
\thispagestyle{empty}
\let\VERBATIM\verbatim
\def\verbatim{%
\def\verbatim@font{\small\ttfamily}%
\VERBATIM}
}

\begin{abstract}

5G network is envisioned to deploy a massive Internet-of-Things (IoTs) with requirements of low-latency, low control overhead and low power. Current 4G network 
is optimized for large bandwidth applications and inefficient to handle short sporadic IoT messages. The challenge here spans multiple layer including the radio access and the network layer. This paper focus on reusing CDMA acess for IoT devices considering event-driven and latency sensitive traffic profile.
We propose a PHY/MAC layer design for CDMA based communication for low power IoT devices. We propose and evaluate coexisting operation of CDMA based IoT network in presence of the exiting LTE network. Our proposed design will integrate IoT traffic with legacy system by minimal modification at the edge network, essentially eNodeB. 
\eat{
{\color{magenta}It needs modification only at interface at edge network component eNodeB to be able to process both LTE data and CDMA IoT packets.}}
We show that the underlay CDMA IoT network meets IoT data traffic requirements with minimal degradation ($3\%$) in the LTE throughput. We also implement the proposed design using Software Defined Radios and show the viability of the proposal under different network scenarios.
\end{abstract}

\eat{

}


\section{Introduction}
With exponential growth of IoT devices \cite{CiscoVNI_Feb16}, the 5G network will experience a variety of traffic patterns not prevalent in earlier 4G systems.
IoT devices often transmits short sporadic messages, which are not well suited to the high data traffic and connection-oriented modes associated with legacy 3GPP networks resulting in high service latency and excessive control overhead. 
In order to access the network, a User Equipment (UE) has to follow attachment, authentication and bearer establishment procedure which accounts for $30\%$ of control plane signaling overhead.
Furthermore, a UE goes into the Idle state if it has been inactive for more than 10 seconds (applicable for many targeted IoT applications) and the UE needs to re-establish the bearer for the next transmission. Latency for the Idle to connected state is $\sim 60$ ms\cite{Singhal2010_latencyAna}.
With the dense deployment of IoT devices, current 4G network will be extensively overwhelmed by the surge in both traffic and control plane signaling load.
The 5G network needs the design provision to accommodate heterogeneous IoT applications at very high scale with low latency and low control overhead across both the radio access network and core network.

The goal is to operate in the same band as current LTE, thus not requiring any separate channel allocation, 
and is backward compatible with the 4G network\cite{Mathur2016_crossLayer}.
The 3GPP proposed NB-IoT radio access technology leverages LTE design extensively, thus making it compatible with existing LTE operation\cite{Wang2016_NBIoTPrimer}. But with the massive deployment of NB-IoT devices, overall network mainly come across with shortcomings of (1) increase in controlling signal overhead and (2) scheduling/sharing spectrum band among NB-IoT and LTE devices. This would lead to diminishing throughput performance of LTE due to decrease in usable data channel for LTE.  

This motivates us to propose CDMA-based low power IoT  transmission for the simultaneous channel access of IoT device along with LTE devices where IoT devices can operate at low Signal-to-Noise-Interference-Ratio (SINR) which is sufficient for low bit-rate data transmission\cite{mathur2017_CDMAIoTDemo}. The purpose of this paper is to present an analytical study of CDMA-based IoT system along with its coexistence with LTE and a proof-of-concept using software-defined radio (SDR) platform. The contributions of the paper are following:
\begin{enumerate}
\item We propose an underlay CDMA-based MAC and Physical layer IoT communication in presence of existing OFDM based system, where minimal changes are required at mobile edge node without any modification in mobile core network.
\vspace{-0.12cm}
\item We performed analytical capacity estimation of CDMA-based IoT network for standalone operation.
\vspace{-0.12cm}
\item We also performed analytical modeling to characterize coexistence performance of LTE and CDMA underlay IoT network and capacity comparison of CDMA underlay network with existing IoT protocol (NB-IoT) with respect to practical IoT data traffic requirements.
\vspace{-0.12cm}
\item Finally, we implemented the proposed design using Software Defined Radios (SDRs) and performed practical experimentation to validate underlay CDMA-based IoT network with exemplary test scenarios (1) performance of a single CDMA link, (2) coexistence of LTE and IoT devices
\end{enumerate}

The rest of the paper is organized as follows. We present the proposed CDMA-based IoT system with MAC/PHY cross-layer design details in \S \ref{sec:sysDes}, followed by the analytical capacity estimation of standalone CDMA-based IoT network in \S \ref{sec:AnaCDMA} and analytical model to estimate the coexistence performance of CDMA underlay and LTE network in \S \ref{sec:coex}. \S \ref{sec:implement} and \ref{sec:eval} describe the implementation details and experimental evaluations of CDMA-based IoT system design, respectively. We conclude the paper in \S \ref{sec:conclude}.
\section{CDMA-based IoT Design}
\label{sec:sysDes}

The cellular IoT devices targets applications such as smart utility (gas/water/electric) metering reports, smart agriculture and smart environment. 
We identify system requirements for optimal waveforms for IoT data transmission, choice of CDMA radio access technology with regards to fulfilling these requirements, details of MAC/PHY layer design parameters 
at transmitter (here, IoT device) and receiver (here, eNB).

The main objective at MAC layer is to schedule IoT traffic with the minimum wait time to access the channel. At physical layer, we need to identify the optimal waveform for the\eat{an underlay} transmission operating at low signal-to-interference-plus-noise ratio (SINR) and bit-error-rate (BER) and without causing significant interference to the overlay LTE transmission. Considering these requirements, the CDMA transmission is the favorable choice for IoT services due to (1) it does not require any UE-eNB handshaking (RACH+RRC) contrary to earlier efforts made in the community\cite{Schaich2014_waveform}, (2) Asynchronous CDMA transmission enables decentralized spectrum access at IoT UEs which reduces wait time to get assigned resources from eNB, (3) Low power CDMA IoT transmission can reject narrow band OFDMA LTE interference without causing significant interference to LTE as well, and (4) CDMA-based IoT transmission utilizes the legacy CDMA support available at cellular network.
One of the major challenges in CDMA transmission with LTE-overlay is interference cancellation at the receiver (here, eNB). Therefore, self-interference cancellation may be introduced at eNB in full duplex mode to reduce the interference.

\section{Analytical Study of CDMA-based IoT System}
\label{sec:AnaCDMA}
CDMA has been studied in the literature \cite{proakis2002communication, Gilhousen91_CDMAcapacity, Hara97_multicarrierCDMA} and implemented for 2G/3G cellular system. Also, underlay CDMA has been proposed in cognitive radio domain~\cite{underlayCDMA_2016}. In our study, we take the reference from the available frameworks and validate the feasibility of underlay CDMA for the implementation of sporadic short IoT messages to fulfill requirements of massive deployment, low latency and short-message communication.

\subsection{Background}
In CDMA\cite{proakis2002communication}, information-bearing baseband signal, $s(t)$, is multiplied by the spreading code, $c(t)$. Let us assume that $T_b$ is the bit interval of $s(t)$ with the information rate of $R = 1/T_b$ bits/sec and $T_c$ is the pulse duration of $c(t)$. For such system, CDMA processing gain is given by
\begin{equation}
\label{eq:procGain}
L_c = \frac{T_b}{T_c} = \frac{W}{R_b}
\end{equation}
where $W \approx 1/T_c$ represents the total spread bandwidth. The probability error for the CDMA system using BPSK modulation is expressed as
\begin{equation}
P_b = Q \left( \sqrt{\frac{2 E_b}{N_0}} \right)
\end{equation}
where $Q(x)$ is $Q$-function, $E_b$ is the energy per bit, $N_0$ is the power spectral density of an equivalent broadband interference (generally, over bandwidth $W$) and $E_b / N_0$ is referred as the bit energy-to-noise density ratio.
\eat{
In CDMA\cite{proakis2002communication}, considering transmission of a binary information using binary PSK (BPSK), information-bearing baseband signal is expressed as
\begin{equation}
\label{eq:dataSeq}
s(t) = \sum_{n = -\infty}^{\infty} a_n g_T(t - nT_b)
\end{equation}
where $a_n = \pm 1, -\infty < n < +\infty, T_b$ is the bit interval with the information rate of $R = 1/T_b$ bits/sec, and $g_T(t)$ is a rectangular pulse of duration $T_b$. This signal is multiplied by the spreading code, which can be expressed as
\begin{equation}
c(t) = \sum_{n = -\infty}^{\infty} c_n p_T(t - nT_c)
\end{equation}
where $c_n = \pm 1$ and $p_T(t)$ is a rectangular pulse of duration $T_c$. For this system, CDMA processing gain is given by
\begin{equation}
\label{eq:procGain}
L_c = \frac{T_b}{T_c} = \frac{W}{R_b}
\end{equation}
where $W \approx 1/T_c$ represents the total spread bandwidth. The probability error for the CDMA system using BPSK modulation is expressed as
\begin{equation}
P_b = Q \left( \sqrt{\frac{2 E_b}{N_0}} \right)
\end{equation}
where $Q(x)$ is $Q$-function, $E_b$ is the energy per bit, $N_0$ is the power spectral density of an equivalent broadband interference (generally, over bandwidth $W$) and $E_b / N_0$ is referred as the bit energy-to-noise density ratio.
}

In the proposed CDMA underlay IoT network, we assume all uplink signals (IoT device-to-eNB) are received at the same power level $S$. For $N$ IoT UEs, signal-to-interference-plus-noise-ratio (SINR) is given by\cite{Gilhousen91_CDMAcapacity}
\begin{equation}
\rm{SINR} = \frac{S}{(N-1)S + \eta}
\end{equation}
and $E_b / N_0$ can be obtained as
\begin{equation}
E_b / N_0 = \frac{S/R_b}{(N-1)S/W + \eta/W} = \frac{W/R_b}{(N-1) + \eta/S}
\end{equation}
For parameters $E_b / N_0$, required for the adequate performance of demodulation and decoding, and $S$, the capacity of CDMA underlay network is
\begin{equation}
\label{eq:numUsers}
N = 1 + \frac{W/R_b}{E_b / N_0} - \frac{\eta}{S}
\end{equation}

\subsection{Capacity of CDMA IoT Network}
\begin{figure}[t]
\vspace{-0.5cm}
\begin{center}
\epsfig{figure= 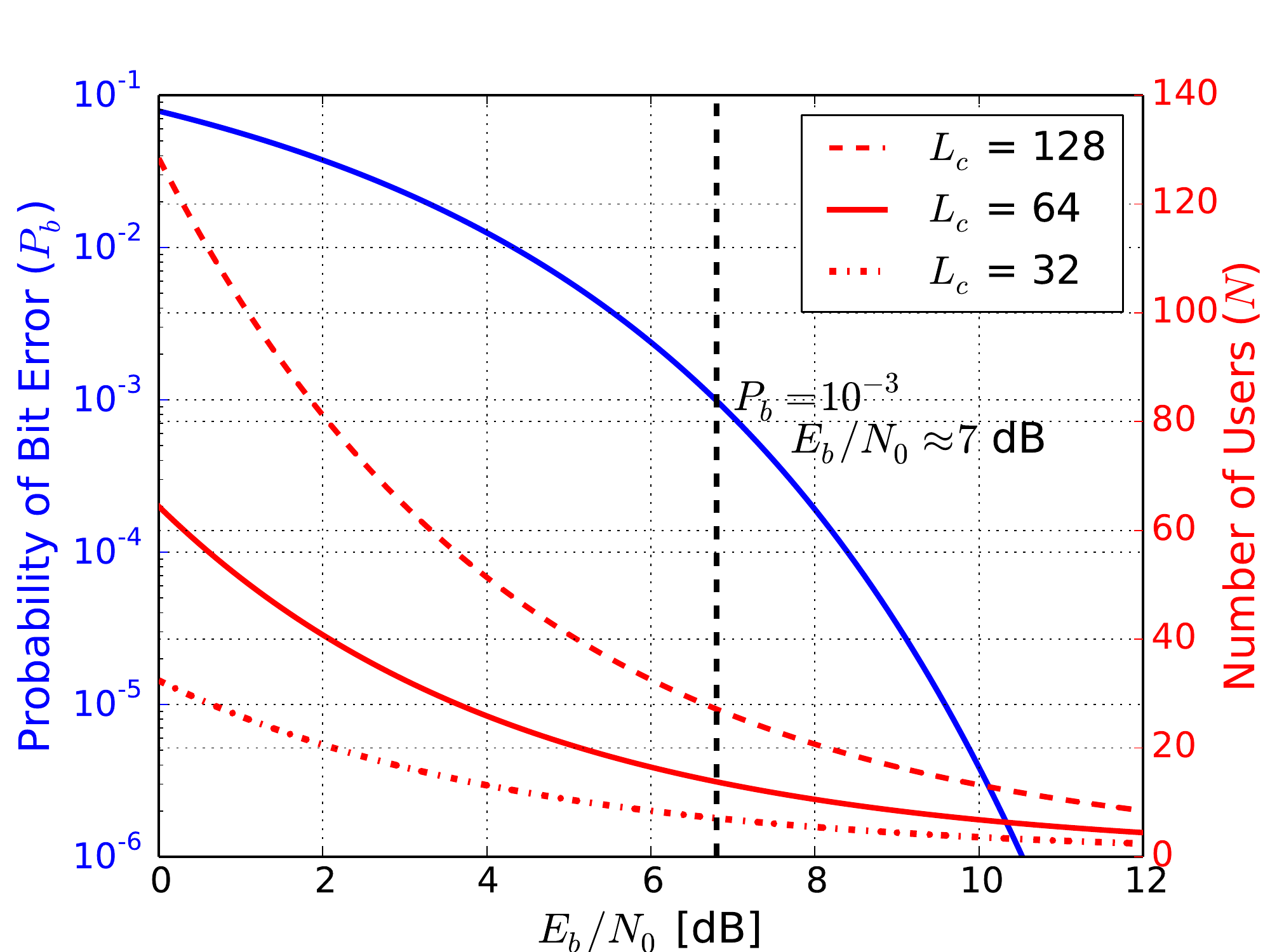,width=2.65in}
\end{center}
\caption{Bit error rate ($P_b$) and capacity of CDMA IoT network ($N$) as a function of $E_b / N_0$ and processing gain ($L_c$) for received power $S$ at distance = 600 m}
\label{fig:numUsers_ebno7}
\vspace{-0.5cm}
\end{figure}

We evaluate the capacity of the proposed CDMA IoT network for typical cellular deployment parameters (given in Table \ref{tab:cdma_para}). As shown in Fig.~\ref{fig:numUsers_ebno7}, the capacity of CDMA IoT network, $N$, can be augmented by increasing processing gain (higher system complexity) or by compromising system performance in terms of bit error rate (BER), $P_b$. Assuming the BER system constraint $P_b = 10^{-3}$ (e.g. required $P_b$ or better for digital voice transmission), the required $E_b/N_0$ is approximately 7 dB at $L_c = 64$. In a single CDMA channel, it can accommodate $N = 13$ IoT UEs simultaneously without affecting each other's performance. The total CDMA network capacity, $N_{\max}$ can be obtained as (Capacity of a single CDMA channel * No. of CDMA channels).
As an example, for CDMA channel with bandwidth 1 MHz and $N=$ 13 and 20 MHz spectrum band (similar to LTE band), maximum CDMA IoT network capacity is $260$. Few main observations which affect the capacity of the CDMA IoT network are discussed next.
\eat{
\begin{equation}
\begin{aligned}
N_{\max} & = \text{Capacity of a single CDMA channel}\\
	&  \qquad \text{* No. of CDMA channels}
\end{aligned}
\end{equation}
}

\begin{table}[t]
\centering
\caption{Parameters for CDMA-based IoT network}
\label{tab:cdma_para}
\begin{tabular}{| l | l |}
\hline
\textbf{System Parameter} & \textbf{Specification}\\
\hline
CDMA bandwidth & 1 MHz\\
\hline
Modulation & BPSK\\
\hline
Antenna configuration& UE: 1T1R\\ 
\hline
Operating frequency & 2620.5 MHz\\
\hline
UE transmit power & 23 dBm\\
\hline
Thermal noise density & -174 dBm/Hz\\
\hline
Cell site sector radius & 600 m\\
\hline
Path loss model (for & 36.7 $\log_{10}$(dist[m]) + 22.7\\
2GHz $<$ freq $<$ 6GHz) & + 26 $\log_{10}$(frq [GHz])\\
\hline
\end{tabular}
\vspace{-5pt}
\end{table}

\subsubsection{No requirement of guard bands}
CDMA IoT network can avoid guard band between CDMA channels by using non-overlapping data spreading codes. For example, 64-bit Hadamard code forms 64 spreading code sequences. For $N=13$, 4 groups of non-overlapping sequences are formed. In this case, maximum channel separation is 4 MHz for $W = 1$ MHz minimizing the interference between adjacent channels.

\subsubsection{Robust performance in multi-cell scenario}
In the conservative approach of limited spectrum band with reuse factor 1, interference between CDMA UEs in adjacent cells can be mitigated by CDMA code and channel diversity. For example, for 20 CDMA channels with $W = 1$ MHz in 20 MHz band and 4 non-overlapping spreading sequence groups, code-channel diversity has a reuse factor of 80. This assures that UEs in adjacent cells with the same channel and spreading code are at least apart by inter-eNB distance causing negligible interference to each other.

\section{Coexistence of CDMA Underlay IoT and LTE Network}
\label{sec:coex}
The CDMA IoT network is proposed to coexist with the existing LTE network in the same spectrum band. In this section, we propose an analytical model to evaluate coexistence performance of CDMA underlay IoT and LTE network. We also evaluate the performance of CDMA network to suffice the data demand with regards to a practical IoT UE deployment and data traffic model. We focus an uplink scenario only.

\subsection{Coexistence Analytical Model}
Let us assume that $N$ CDMA underlay IoT devices per CDMA channel are uniformly distributed over a hexagonal mobile site with radius $r_d$. Assuming IoT UEs employ power control such that eNB receive an equal power $P_r^C$ from UEs. Assuming the worst-case IoT UE is located at distance $r_d$ from eNB, $P_t^C$ is the maximum transmit power and $PL^C(r_d, f)$ is the path loss at the distance $r_d$ and frequency $f$, 
$P_r^C$ is given as
\begin{equation}
P_r^C = \frac{P_t^C}{PL^C(r_d, f)},
\end{equation}

For LTE network, we assume that $M$ UEs are uniformly distributed over the mobile site, $R$ resource blocks are available for LTE uplink operation and each LTE UE get $M/R$ contiguous resource blocks. Assuming that all LTE UEs are transmitting at maximum power $P_t^L$, the total average power, $Q_{R_c}^L$, received at the eNB from LTE UE  over CDMA channel bandwidth equivalent to $R^C$ resource blocks is \cite{Grieco1994_CDMAunderlay}
\begin{equation}
Q_{R_c}^L = 3\beta \frac{M R_c}{R} \frac{P_t^L}{PL^C(r_d, f)} 
\end{equation}
where $\beta$ is the LTE channel occupancy factor and ($P_t^L M/R$) is the transmission power per LTE resource block.

CDMA $E_b/N_0$ at the eNB is given by
\begin{equation}
E_b/N_0 = \frac{L_c P_r^C}{N P_r^C + Q_{R_c}^L + \eta}
\end{equation}
and CDMA-based IoT capacity can be obtained as
\begin{equation}
N = 1 + \frac{L_c}{E_b/N_0} - \frac{Q_{R_c}^L + \eta}{P_r^C}
\end{equation}

Power ($P_{r,W}^L, P_{r,A}^L$) received by eNB from the worst-case and average-case LTE UEs located at the cell boundary $r_d$ and at average distance $r_d/2$ is given by
\begin{equation}
P_{r,W}^L = \frac{P_t^L}{PL^L(r_d, f)}, P_{r,A}^L = \frac{P_t^L}{PL^L(r_d/2, f)}.
\end{equation}

Uplink LTE SINR ($\Gamma^L_W, \Gamma^L_A$) can be obtained as
\begin{equation}
\Gamma^L_W = \frac{K. P_{r,W}^L}{N P_r^C + \eta}, \Gamma^L_A = \frac{K. P_{r,A}^L}{N P_r^C + \eta},
\end{equation}
respectively and $K = R^C M / R$ is an overlapping bandwidth factor. LTE throughput can be computed as \cite{Sagari2015_LTEUWiFi}
\begin{equation}
T^L_x = a W_L \log_2 (1 + b \Gamma^L_x), \quad x = \{W, A\}
\end{equation}
where $W_L$ is the LTE channel bandwidth, $a$ is a factor associated to the gap between Shannon and actual capacity and $b$ is the LTE bandwidth efficiency.

\begin{figure}[t]
\begin{center}
\epsfig{figure= 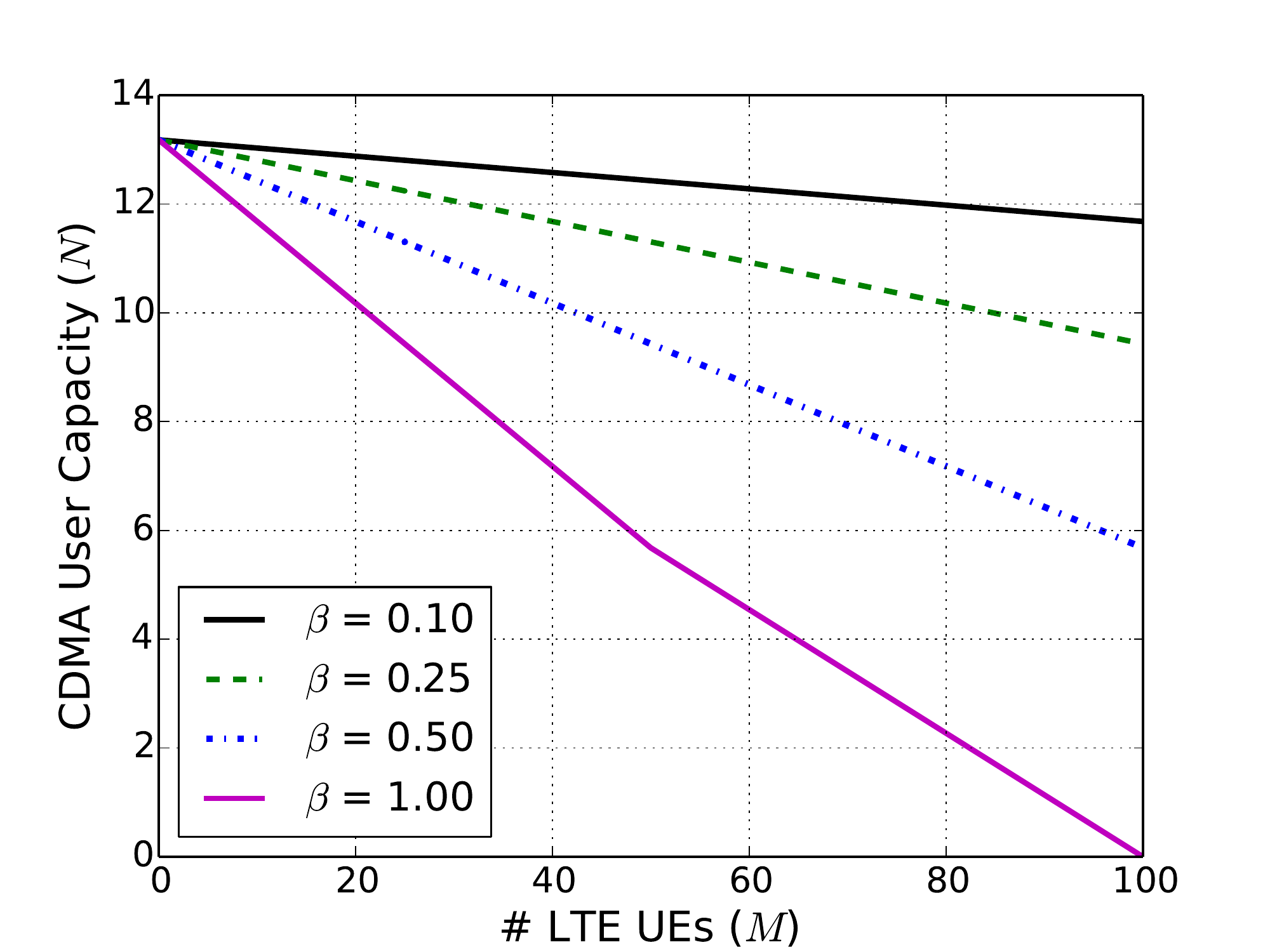,width=2.65in}
\end{center}
\caption{Capacity of CDMA-based IoT Network as a function of no. of LTE UEs and LTE channel occupancy ($\beta$) when required CDMA $E_b/N_0$ = 7 dB, $W = 1$MHz.
}
\label{fig:coex_CDMA_capacity}
\vspace{-0.5cm}
\end{figure}

Fig.~\ref{fig:coex_CDMA_capacity} shows the capacity of CDMA underlay IoT network as a function of number of LTE UEs $M$ and LTE channel occupancy $\beta$. It shows that for the worst-case, when LTE data requirement is high (e.g. $M=100, \beta = 1$), the simultaneous CDMA underlay traffic is not possible and the IoT data traffic needs a dedicated channel assignment. For the average case, the LTE network can share the spectrum with multiple IoT UEs.

\begin{figure}[t]
\begin{center}
\epsfig{figure= 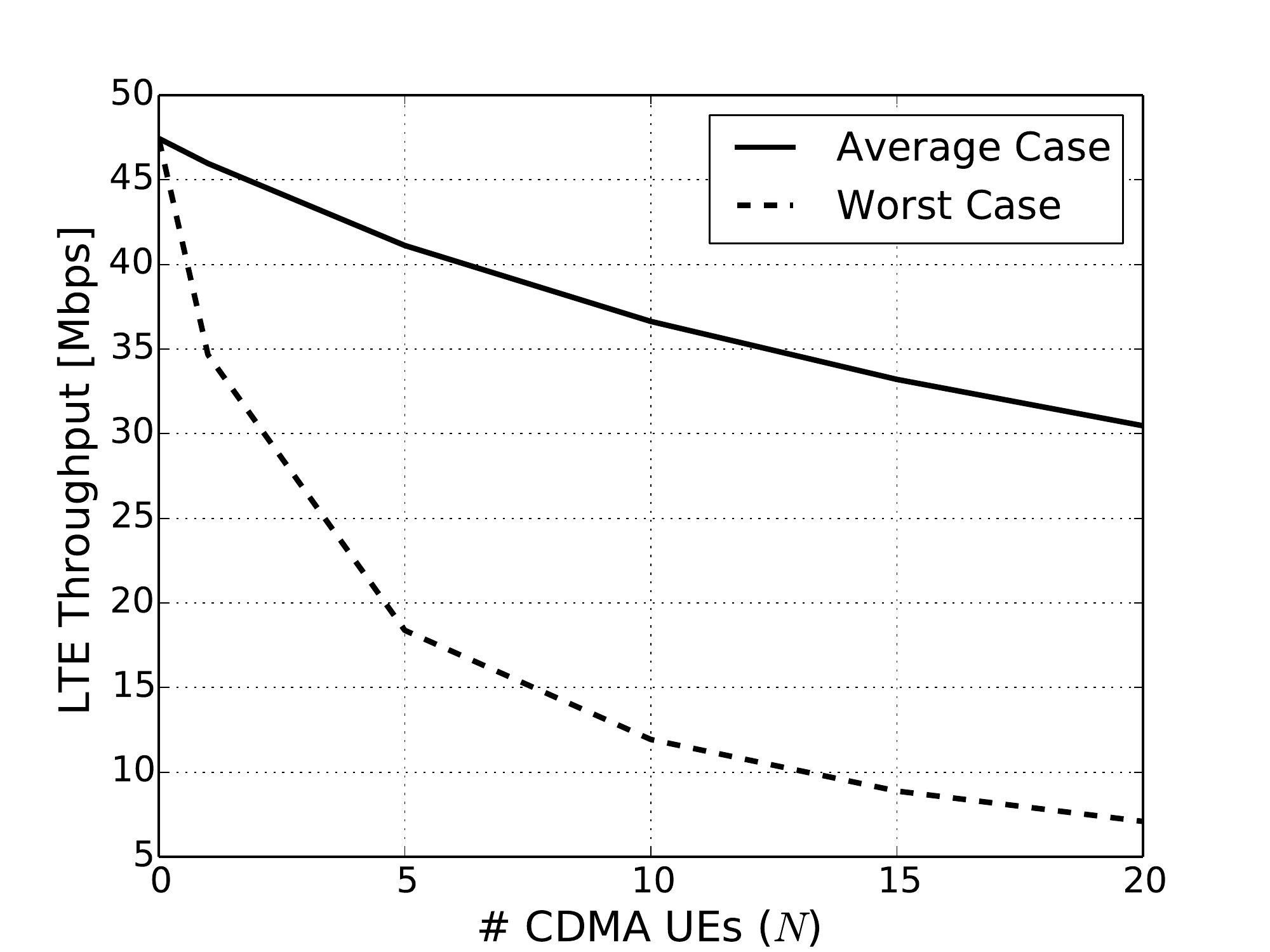,width=2.65in}
\end{center}
\caption{LTE throughput for a single average and worst-case user as a function of number of CDMA UEs
}
\label{fig:ltethru_cdmaCoex}
\vspace{-0.5cm}
\end{figure}
Fig.~\ref{fig:ltethru_cdmaCoex} shows the average and worst case LTE throughput as a function of number of CDMA-based IoT UEs, $N$. As $N$ increases the LTE throughput decreases but along with gaining the IoT UE capacity. For the average case, the LTE throughput decreases by $3\%$ and $12\%$ when $N$ is 5 and 10, respectively. The analysis assumes that CDMA and LTE UEs transmit at the maximum power. But both CDMA and LTE employ power control based on CDMA/LTE co-channel interference causing lesser interference to other uplink transmission. 
Thus, this plot represents the conservative (lower bound) LTE throughput coexisting with CDMA underlay IoTs.

\subsection{Practical IoT Data Demand vs Capacity of CDMA-underlay Network}
In urban scenario, the expected cellular IoT device density per cell site sector is $N_{MS} = 52,547$ device\cite{3gpp_nbIoT}. 
We adopt the IoT data traffic profile defined in \cite{3gpp_nbIoT} and compare the capacity of the proposed CDMA-underlay IoT with NB-IoT with respect to meeting IoT data requirement in urban deployment. IoT UEs report periodic updates and exception reports.
The ideal capacity requirement from $N_{MS}$ IoT UEs per site sector per day is calculated by considering distribution of periodicity and payload size of IoT data (see Table \ref{tab:IoT_Syspara}).   

\begin{table}[t]
\centering
\caption{IoT system parameters}
\label{tab:IoT_Syspara}
\begin{tabular}{| l | l |}
\hline
\textbf{Parameter} & \textbf{Value}\\
\hline
\multicolumn{2}{|l|}{\textbf{IoT Data Traffic Model}}\\
\hline
IoT Application & Pareto distribution: $\alpha = 2.5$ (shape\\
payload size ($PL$) & parameter), min $PL$ = 20 bytes,\\
distribution&max $PL$ = 200 bytes (for\\
& $PL > 200$ bytes, $PL = 200$ bytes).\\
\hline
Periodicity split & 1 day (0.4), 2 hours (0.4),\\
for IoT data & 1 hour (0.15), 30 minutes (0.05)\\
\hline
\multicolumn{2}{|l|}{\textbf{CDMA-based IoT Network Parameters }}\\
\hline
Packet overhead & 65 Bytes (no IP compression)\\
\hline
Max payload size & 20 Bytes\\
\hline
Processing gain & 64 with $W = 1$ MHz\\
\hline
\end{tabular}
\end{table} 

\begin{figure}[t]
\begin{center}
\epsfig{figure= 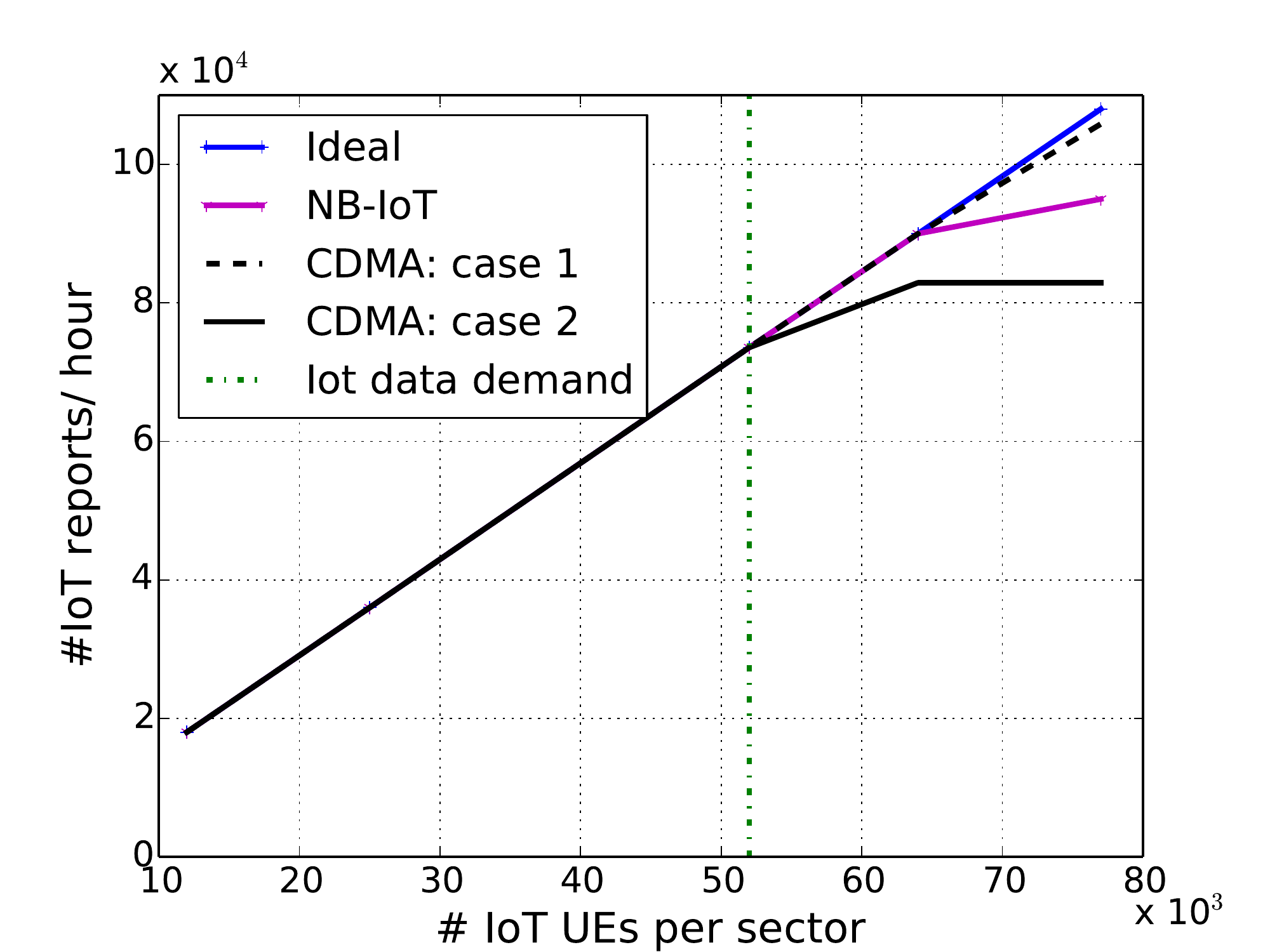,width=2.65in}
\end{center}
\caption{Comparison of capacity of CDMA-based IoT network with NB-IoT, CDMA case 1: $N$ = 5, data repetition factor ($\delta$)=2, CDMA case 2: $N$ = 5, $\delta$=3}
\label{fig:nbIoT_CDMA_capa}
\vspace{-0.5cm}
\end{figure}

Fig.~\ref{fig:nbIoT_CDMA_capa} compares the ideal capacity requirement with capacity of NB-IoT (existing protocol) \cite{Wang2016_NBIoTPrimer} and CDMA underlay IoT.
For CDMA based IoT, simultaneous operation of $N = 5$ IoT UEs are allowed and simulations are run for data repetition factors $\{2,3\}$ and parameters given in Table~\ref{tab:IoT_Syspara}. There is no significant difference in ideal capacity and capacity achieved CDMA-based IoT and NB-IoT for $N_{MS} = 52600$. It implies that CDMA underlay IoT with $N = 5$ satisfies IoT traffic demand affecting LTE throughput only up to $3\%$.
\vspace{20pt}
\section{IoT System Implementation}
\label{sec:implement}

\begin{figure}[t]
\begin{center}
\epsfig{figure= 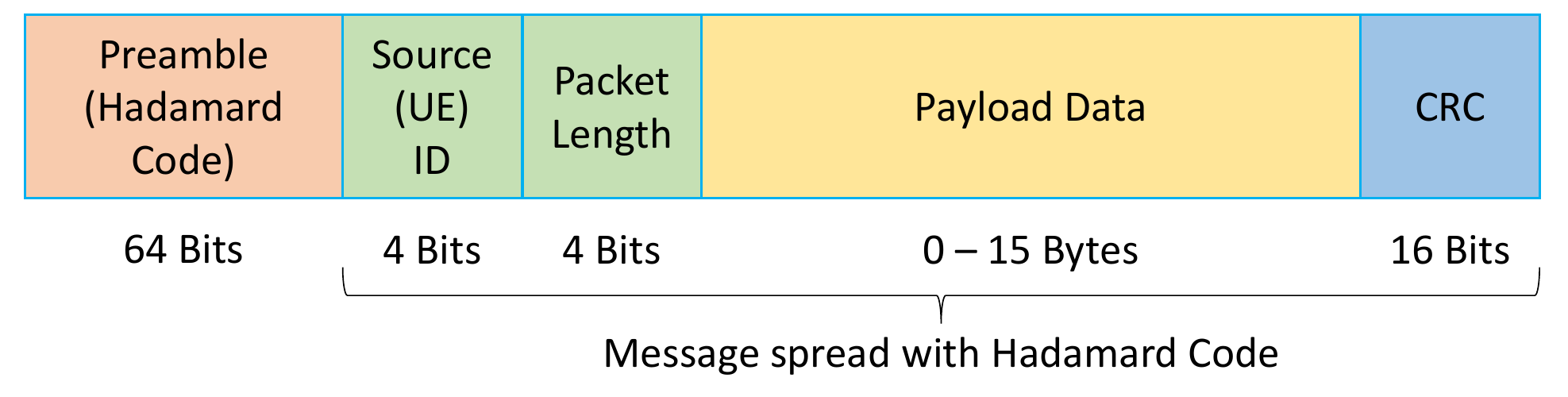,width=\linewidth}
\end{center}
\caption{Packet format of CDMA data transmission}
\label{fig:ProtoSynTax}
\vspace{-0.5cm}
\end{figure}

We have prototyped CDMA-based IoT transmission using a software-defined radio (SDR) platform using GNU radio and  Universal Software Radio Peripheral (USRP). USRP Hardware Drivers (UHD) are used to transmit/receive samples to/from the USRP. The CDMA transmitter and receiver code is developed on top of UHD in C++ and C, respectively, to process and decode CDMA packets in real time. Intel(R) Core i7 4th Generation CPU ($@$3.60 GHz) machines are used in performance mode. 
We adopt an example packet format instance as shown in Fig.~\ref{fig:ProtoSynTax} for the uplink IoT data transmission. The CDMA MAC frame consists of source address, payload and cyclic Redundancy Check (CRC) to check errors. We can send variable size payload of 0-15 Bytes which is typical IoT traffic profile emanating from sensing devices\cite{LPWA2016}. The MAC frame is spreaded/despreaded with the 64-bit Hadamard code assigning unique codes to each UE. Data is modulated/ demodulated using BPSK to operate on low-SINR. Packet preamble contains a unique Hadamard code to detect the start of the packet at the receiver by correlating signal with the known code. The CDMA channel uses 1 MHz bandwidth with data rate $\sim$10 - 15 Kbps.

Developing CDMA receiver is particularly challenging to detect CDMA packet considering random wireless channel and critical time constraints of real-time signal processing. For example, to detect beginning of the packet, we perform cross-correlation of a known 64-bit preamble and 10,000 received samples (equivalent to size of a packet) at every instance. Each instance takes processing time of 160 $\mu$s including 100 $\mu$s for reading samples at sampling rate of 1 MSps and 60 $\mu$s for running correlation function. We choose sampling rate 1 MSps to avoid overflowing of samples at receiver which causes due to higher sampling rate. This parameter also restrict the maximum achievable data rate for the CDMA transmission. 
Furthermore, a packet is detected if the peak value of the cross-correlation output is greater than certain threshold which is a function of SINR. Here the choice of the threshold becomes critical. If it very low, then there is significant false packets detection which eventually gets discarded while checking the packet's CRC. At higher threshold, the packets gets missed in the detection function.

\section{Evaluation}
\label{sec:eval}

We evaluate our prototype for the following scenarios- (1) standalone operation of CDMA IoT transmission and, (2) coexistence of CDMA IoT and LTE transmissions. IoT nodes are realized using USRP series N210 and/or B210. LTE transmission is enabled using openairinterface (OAI) where OAI is a PC-hosted open sourced SDR\eat{PHY layer prototyping} platform \cite{Nikaein2014_OAI}.
The LTE UE is connected to the eNB using FDD mode, 5 MHz bandwidth and transmission mode 1 (SISO). 

\begin{figure}[t]
\begin{center}
\vspace{0.2in} \hspace{0.35in}
\epsfig{figure= 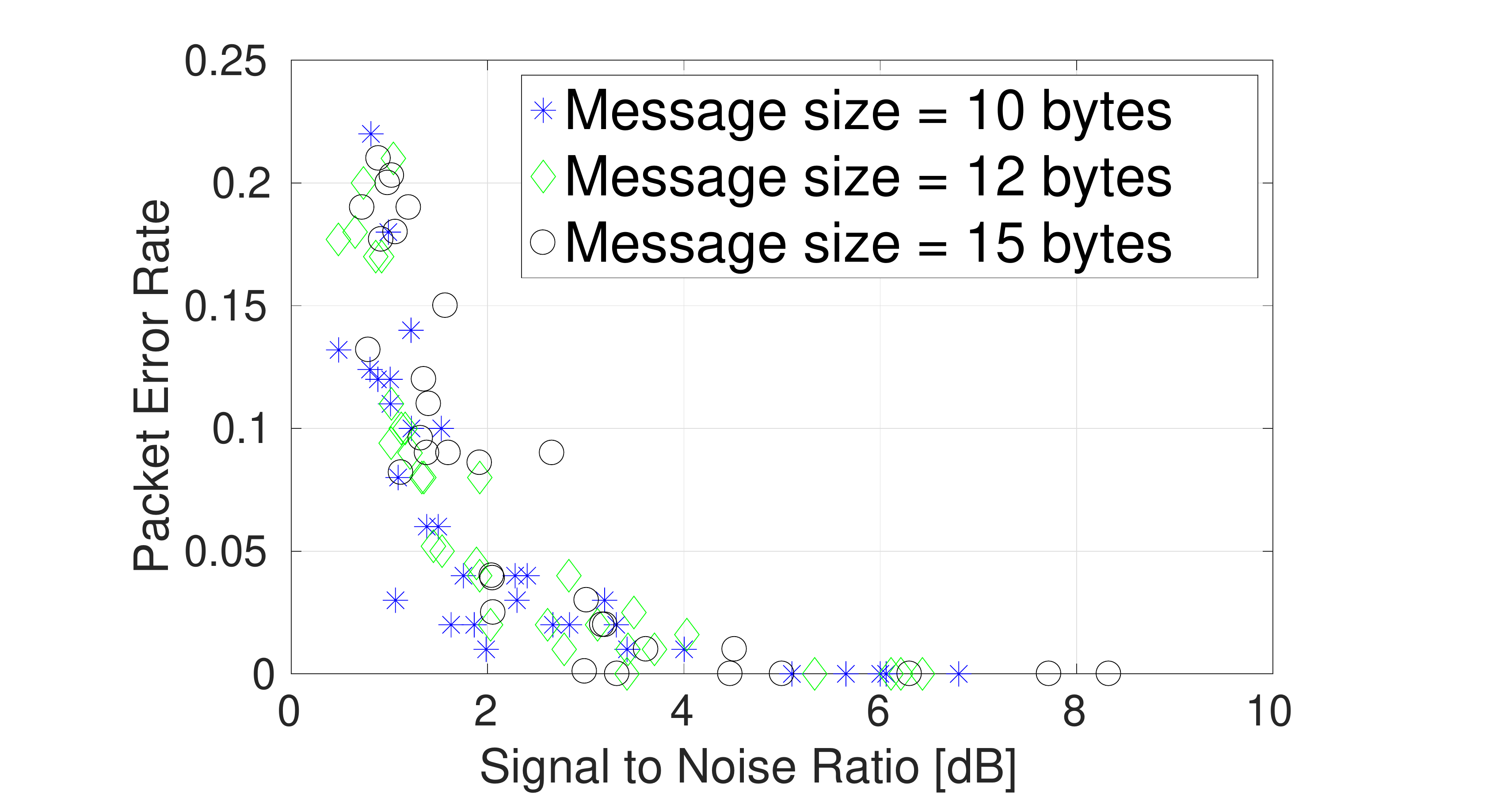,width=3.05in}
\end{center}
\caption{Packet Error Rate of a CDMA-based IoT transmission as a function of the Signal-to-Noise Ratio}
\label{fig:CDMA_PERvsSNR}
\vspace{-0.5cm}
\end{figure}

\subsection{Standalone operation of CDMA IoT}
\label{sec:standCDMA}

\eat{ Fig.~\ref{fig:IoT_eNB_twoUEs} shows the experimental setup on ORBIT testbed for evaluating the packet error rate (PER) performance of a single uplink CDMA IoT transmission as a function of Signal-to-Noise-Ratio (SNR) (see Fig.~\ref{fig:CDMA_PERvsSNR}). } 
We have chosen two nodes in ORBIT grid as the IoT eNodeB and IoT Client, where the distance between the two chosen nodes was 17m. Figure \ref{fig:CDMA_PERvsSNR} shows the packet error rate (PER) of a single uplink CDMA IoT transmission as a function of Signal-to-Noise-Ratio (SNR).
In our experiment, we varied packet payload size, $PL$, as $\{10, 12,15\}$ Bytes. For each $PL$, packets are transmitted at the interval of 60 milliseconds and, with the constant receiver gain, transmitter gain is adjusted to vary SNR between 0 to 8 dB. We observed that PER is between 0.22 to 0 for SNR range from 0 to 4.5 dB and PER is always zero for SNR above 4.5 dB. Also, PER does not get affected significantly due $PL$ values considered in the experiment. Through these set of experiments, we show that our prototyped CDMA IoT transmission can transmit data with low PER with significantly low SNR value along with varying packet payload size.
\eat{
The scenario of two IoT UEs connected to the single eNB poses the challenge of increased signal processing complexity at the receiver (eNB) where each UE spreads the message with separate Hadamard code and data transmission is asynchronous. This scenario is evaluated against the metrics used in the earlier scenario. This case could further be extended for multiple ($>$ 2) UEs. 
}
\begin{figure}[t]
\begin{center}
\epsfig{figure= 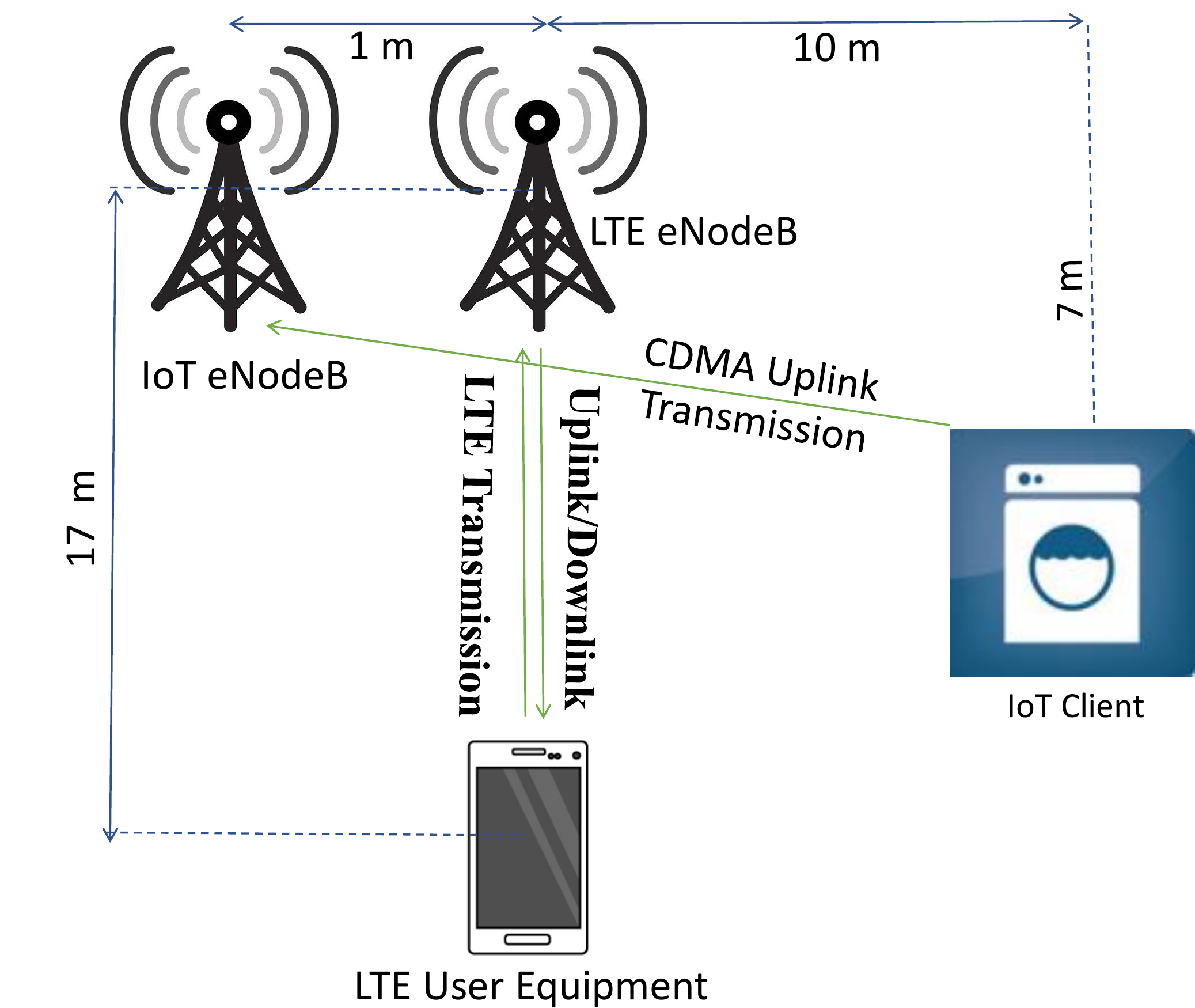,width=2.4in}
\end{center}
\caption{Setup for coexistence of IoT link and LTE using CDMA-underlay implementation, OAI and USRP}
\label{fig:IoT_LTE_coex}
\vspace{-0.5cm}
\end{figure}

\subsection{Coexistence of CDMA IoT and LTE}
\begin{figure}[t]
\begin{center}
\epsfig{figure= 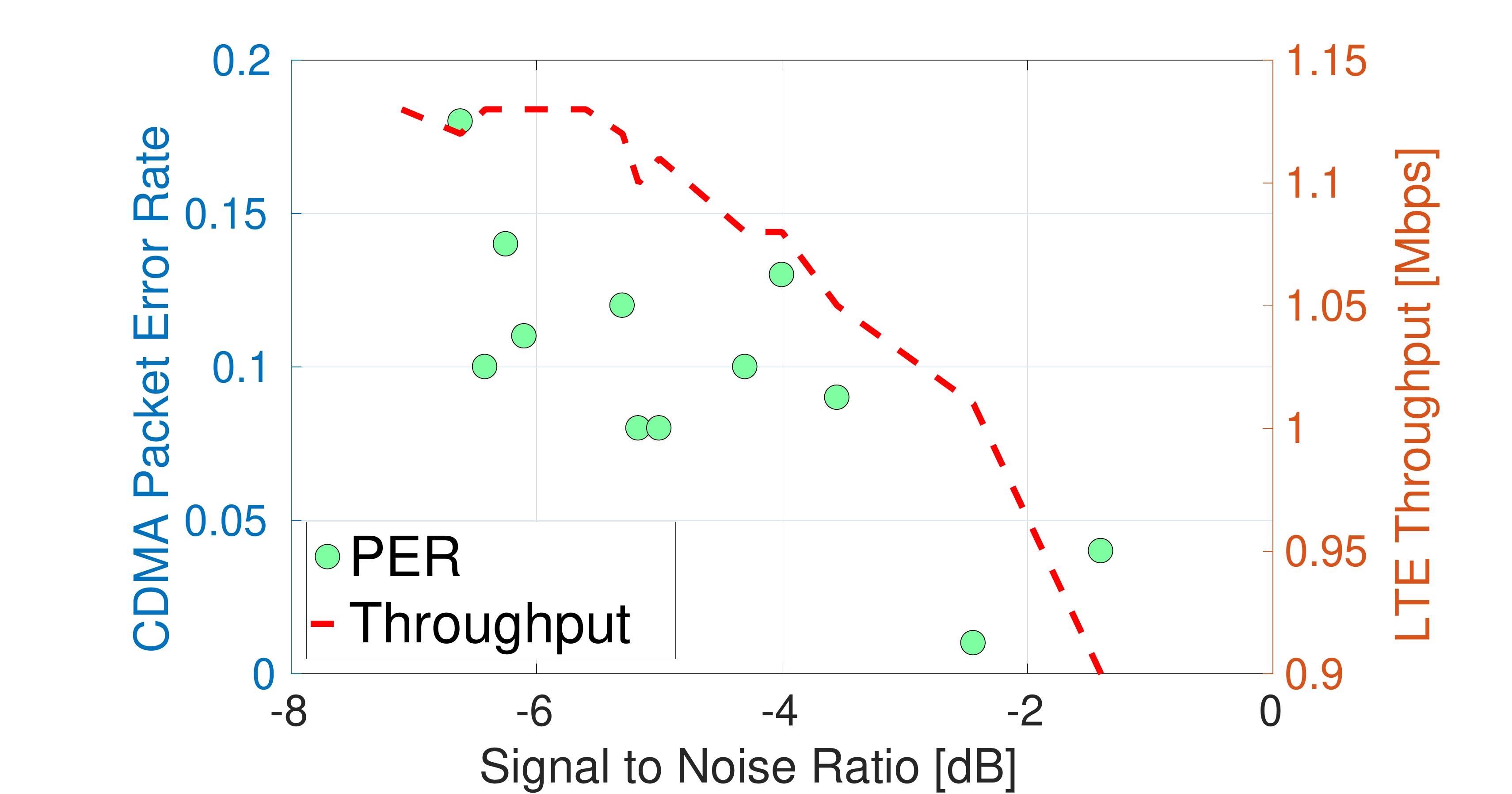,width=3.05in}
\end{center}
\caption{LTE Throughput and Packet Error Rate of the CDMA as a function of the CDMA SNR}
\label{fig:LTEwithCDMA}
\vspace{-0.3cm}
\end{figure}

Experiment setup shown in Fig.~\ref{fig:IoT_LTE_coex} targets to evaluate performance of the shared band operation of LTE and CDMA-underlay IoT transmissions. In this setup, we have separate eNBs for CDMA IoT and LTE UEs. We are currently integrating IoT eNB and OAI/LTE eNB into one unified eNB. In our experiments, for CDMA IoT transmission, packet payload and interval between packet transmission is kept constant at 15 Bytes and 60 millisecond, respectively. CDMA IoT SNR is varied as mentioned in Section \ref{sec:standCDMA}. LTE UE transmits UDP traffic using iperf. LTE is not completely loaded and achieves 1.15 Mbps throughput without any CDMA transmission. As shown in Fig.~\ref{fig:LTEwithCDMA}, with simultaneous CDMA and LTE transmission, CDMA average SNR is observed to be in the range -7.1 dB to -1.4 dB and PER is in the range of 0.01 to 0.22 (due to not fully loaded LTE channel). As CDMA SNR increases, LTE throughput decreases maximum up to $20\%$. This scenario provides insight on design details of CDMA-underlay IoT protocol considering OFDMA structure of LTE.  

\section{Conclusion}
\label{sec:conclude}
With the massive deployment of IoT devices, the current 4G network will be overwhelmed by the surge in control plane signaling overhead and network latency.
This motivated us to design a CDMA-based cross-layer PHY/MAC protocol for IoT devices. The proposed IoT system coexists with in-band LTE operation and allows uplink sporadic IoT data transmission without a need of resource allocation from the LTE network.
We implemented the proposed protocol using software-defined radio platform for exemplary scenarios.
\bibliographystyle{unsrt}
\bibliography{ref}

\end{document}